\documentclass[a4paper]{jpconf}
\usepackage{graphicx}
\usepackage{xspace}
\usepackage{overpic}
\usepackage{ifthen} % for conditional statements

\newboolean{uprightparticles}
\setboolean{uprightparticles}{false} %True for upright particle symbols

\usepackage{xspace} 
\usepackage{upgreek}

\def\MagUp {\mbox{\em Mag\kern -0.05em Up}\xspace}

\ifthenelse{\boolean{uprightparticles}}%
{

 \def\PDelta      {\ensuremath{\Delta}\xspace}                 
 \def\PXi      {\ensuremath{\Xi}\xspace}                 
 \def\PLambda      {\ensuremath{\Lambda}\xspace}                 
 \def\PSigma      {\ensuremath{\Sigma}\xspace}                 
 \def\POmega      {\ensuremath{\Omega}\xspace}                 
 \def\PUpsilon      {\ensuremath{\Upsilon}\xspace}                 
 
 \def\PB      {\ensuremath{\mathrm{B}}\xspace}                 
                  
 \def\PD      {\ensuremath{\mathrm{D}}\xspace}

 \def\PK      {\ensuremath{\mathrm{K}}\xspace}

 \def\Pi      {\ensuremath{\mathrm{i}}\xspace}

 \def\Pp      {\ensuremath{\mathrm{p}}\xspace}

}
{

 \mathchardef\PDelta="7101
 \mathchardef\PXi="7104
 \mathchardef\PLambda="7103
 \mathchardef\PSigma="7106
 \mathchardef\POmega="710A
 \mathchardef\PUpsilon="7107
                  
 \def\PB      {\ensuremath{B}\xspace}                 
                  
 \def\PD      {\ensuremath{D}\xspace}

 \def\PK      {\ensuremath{K}\xspace}

 \def\Pi      {\ensuremath{i}\xspace}

 \def\Pp      {\ensuremath{p}\xspace}

}

\makeatletter
\ifcase \@ptsize \relax% 10pt
  \newcommand{\miniscule}{\@setfontsize\miniscule{4}{5}}% \tiny: 5/6
\or% 11pt
  \newcommand{\miniscule}{\@setfontsize\miniscule{5}{6}}% \tiny: 6/7
\or% 12pt
  \newcommand{\miniscule}{\@setfontsize\miniscule{5}{6}}% \tiny: 6/7
\fi
\makeatother

\DeclareRobustCommand{\optbar}[1]{\shortstack{{\miniscule (\rule[.5ex]{1.25em}{.18mm})}
  \\ [-.7ex] $#1$}}

  \def\Kbar    {{\kern 0.2em\overline{\kern -0.2em \PK}{}}\xspace}

\def\KorKbar    {\kern 0.18em\optbar{\kern -0.18em K}{}\xspace}

  \def\Dbar    {{\kern 0.2em\overline{\kern -0.2em \PD}{}}\xspace}

\def\DorDbar    {\kern 0.18em\optbar{\kern -0.18em D}{}\xspace}

\def\Bbar    {{\ensuremath{\kern 0.18em\overline{\kern -0.18em \PB}{}}}\xspace}

\def\BorBbar    {\kern 0.18em\optbar{\kern -0.18em B}{}\xspace}

  \def\Y#1S{\ensuremath{\PUpsilon{(#1S)}}\xspace}% no space before {...}!

\def\Lbar        {{\ensuremath{\kern 0.1em\overline{\kern -0.1em\PLambda}}}\xspace}
\def\LorLbar    {\kern 0.18em\optbar{\kern -0.18em \PLambda}{}\xspace}

\def\to                 {\ensuremath{\rightarrow}\xspace}

\def\AT#1     {\ensuremath{A_{\mathrm{T}}^{#1}}\xspace}           % 2

\def\C#1      {\ensuremath{\mathcal{C}_{#1}}\xspace}                       % 9
\def\Cp#1     {\ensuremath{\mathcal{C}_{#1}^{'}}\xspace}                    % 7
\def\Ceff#1   {\ensuremath{\mathcal{C}_{#1}^{\mathrm{(eff)}}}\xspace}        % 9  
\def\Cpeff#1  {\ensuremath{\mathcal{C}_{#1}^{'\mathrm{(eff)}}}\xspace}       % 7
\def\Ope#1    {\ensuremath{\mathcal{O}_{#1}}\xspace}                       % 2
\def\Opep#1   {\ensuremath{\mathcal{O}_{#1}^{'}}\xspace}                    % 7

\newcommand{\tev}{\ifthenelse{\boolean{inbibliography}}{\ensuremath{~T\kern -0.05em eV}\xspace}{\ensuremath{\mathrm{\,Te\kern -0.1em V}}}\xspace}
\newcommand{\gev}{\ensuremath{\mathrm{\,Ge\kern -0.1em V}}\xspace}
\newcommand{\mev}{\ensuremath{\mathrm{\,Me\kern -0.1em V}}\xspace}
\newcommand{\kev}{\ensuremath{\mathrm{\,ke\kern -0.1em V}}\xspace}
\newcommand{\ev}{\ensuremath{\mathrm{\,e\kern -0.1em V}}\xspace}
\newcommand{\gevc}{\ensuremath{{\mathrm{\,Ge\kern -0.1em V\!/}c}}\xspace}
\newcommand{\mevc}{\ensuremath{{\mathrm{\,Me\kern -0.1em V\!/}c}}\xspace}
\newcommand{\gevcc}{\ensuremath{{\mathrm{\,Ge\kern -0.1em V\!/}c^2}}\xspace}
\newcommand{\gevgevcccc}{\ensuremath{{\mathrm{\,Ge\kern -0.1em V^2\!/}c^4}}\xspace}
\newcommand{\mevcc}{\ensuremath{{\mathrm{\,Me\kern -0.1em V\!/}c^2}}\xspace}

\def\invfb   {\ensuremath{\mbox{\,fb}^{-1}}\xspace}

\def\ms   {\ensuremath{{\mathrm{ \,ms}}}\xspace}

\def\gsim{{~\raise.15em\hbox{$>$}\kern-.85em
          \lower.35em\hbox{$\sim$}~}\xspace}
\def\lsim{{~\raise.15em\hbox{$<$}\kern-.85em
          \lower.35em\hbox{$\sim$}~}\xspace}

\def\pt         {\mbox{$p_{\mathrm{ T}}$}\xspace}

\def\tell1  {TELL1\xspace}
\def\ukl1   {UKL1\xspace}

 \def\Pp      {\ensuremath{\mathrm{p}}\xspace} 
 \def\PSigma      {\ensuremath{\Sigma}\xspace}

\def\pizero{\ensuremath{\pi^0}\xspace}
\def\ks{\ensuremath{K^0_S}\xspace}
\def\kl{\ensuremath{K^0_L}\xspace}

\def\kpipipi{\ensuremath{K^+ \to \pi^+ \pi^- \pi^+}\xspace}
\def\kpimumu{\ensuremath{K^+ \to \pi^+ \mu^- \mu^+}\xspace}

\def\ksmumu{\ensuremath{K^0_S \to  \mu^- \mu^+}\xspace}
\def\kpimumu{\ensuremath{K^+ \to \pi^+ \mu^- \mu^+}\xspace}

\def\ksmumu{\ensuremath{\ks \to \mu^+\mu^-}\xspace}

\def\kspizmumu{\ensuremath{\ks \to \pi^0 \mu^+\mu^-}\xspace}
\def\klpizmumu{\ensuremath{\kl \to \pi^0 \mu^+\mu^-}\xspace}
\def\kspipiee{\ensuremath{\ks \to \pi^+\pi^- e^+ e^-}\xspace}
\def\ksfourl{\ensuremath{\ks \to \ell^+\ell^- \ell^+ \ell^-}\xspace}
\def\klfourl{\ensuremath{\kl \to \ell^+\ell^- \ell^+ \ell^-}\xspace}
\def\kfourl{\ensuremath{K^0 \to \ell^+\ell^- \ell^+ \ell^-}\xspace}
\def\ksmumuee{\ensuremath{\ks \to \mu^+\mu^- e^+ e^-}\xspace}
\def\kseeee{\ensuremath{\ks \to e^+ e^- e^+ e^-}\xspace}
\def\ksmumumumu{\ensuremath{\ks \to \mu^+\mu^- \mu^+ \mu^-}\xspace}

\def\ksmumu{\ensuremath{K^0_S \to \mu^+ \mu^-}\xspace}

\def\kspizmumu{\ensuremath{K^0_S \to \pi^0 \mu^+ \mu^-}\xspace}
\def\kspipiee{\ensuremath{K^0_S \to \pi^+ \pi^- e^+ e^-}\xspace}
\def\ksmumumumu{\ensuremath{K^0_S \to \mu^+ \mu^- \mu^+ \mu^-}\xspace}
\def\ksmumuee{\ensuremath{K^0_S \to \mu^+ \mu^- e^+ e^-}\xspace}
\def\kseeee{\ensuremath{K^0_S \to e^+ e^- e^+ e^-}\xspace}

\def\kpipipi{\ensuremath{K^+ \to \pi^+ \pi^- \pi^+ }\xspace}

\def\kpimumu{\ensuremath{K^+ \to \pi^+ \mu^- \mu^+ }\xspace}

\def\sigmapmumu{\ensuremath{\PSigma^+ \to \Pp \mu^+ \mu^-}\xspace}

\def\pizero{\ensuremath{\pi^0}\xspace}

\def\kpipipi{\ensuremath{K^+ \to \pi^+ \pi^- \pi^+}\xspace}
\def\kpimumu{\ensuremath{K^+ \to \pi^+ \mu^- \mu^+}\xspace}

\usepackage{lineno}
\begin{document}
\title{Prospects for kaon physics at LHCb}

\author{Francesco Dettori \footnote{On behalf of the LHCb collaboration }}

\address{Universit\`{a} degli Studi di Cagliari and INFN, Cagliari, Italy}

\ead{francesco.dettori@cern.ch}

\begin{abstract}
Despite not being designed for it, the LHCb experiment has given world-leading contributions 
in kaon and hyperon physics. In this contribution I review the prospects for kaon physics at LHCb 
exploiting the already acquired data and the current and future Upgrade scenarios. 
\end{abstract}

\begin{figure}
\begin{overpic}[width = 0.68 \textwidth  ]{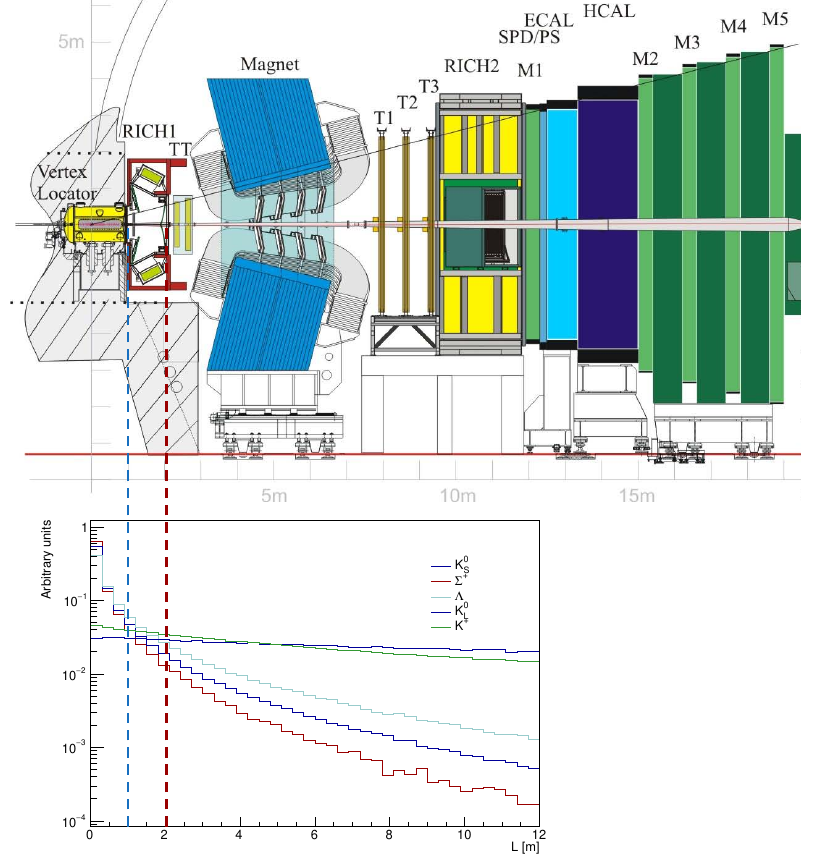}
\put(20,95){\small (b)}
\put(20,30){\small (a)}
\end{overpic}
\begin{overpic}[width = 0.32\textwidth]{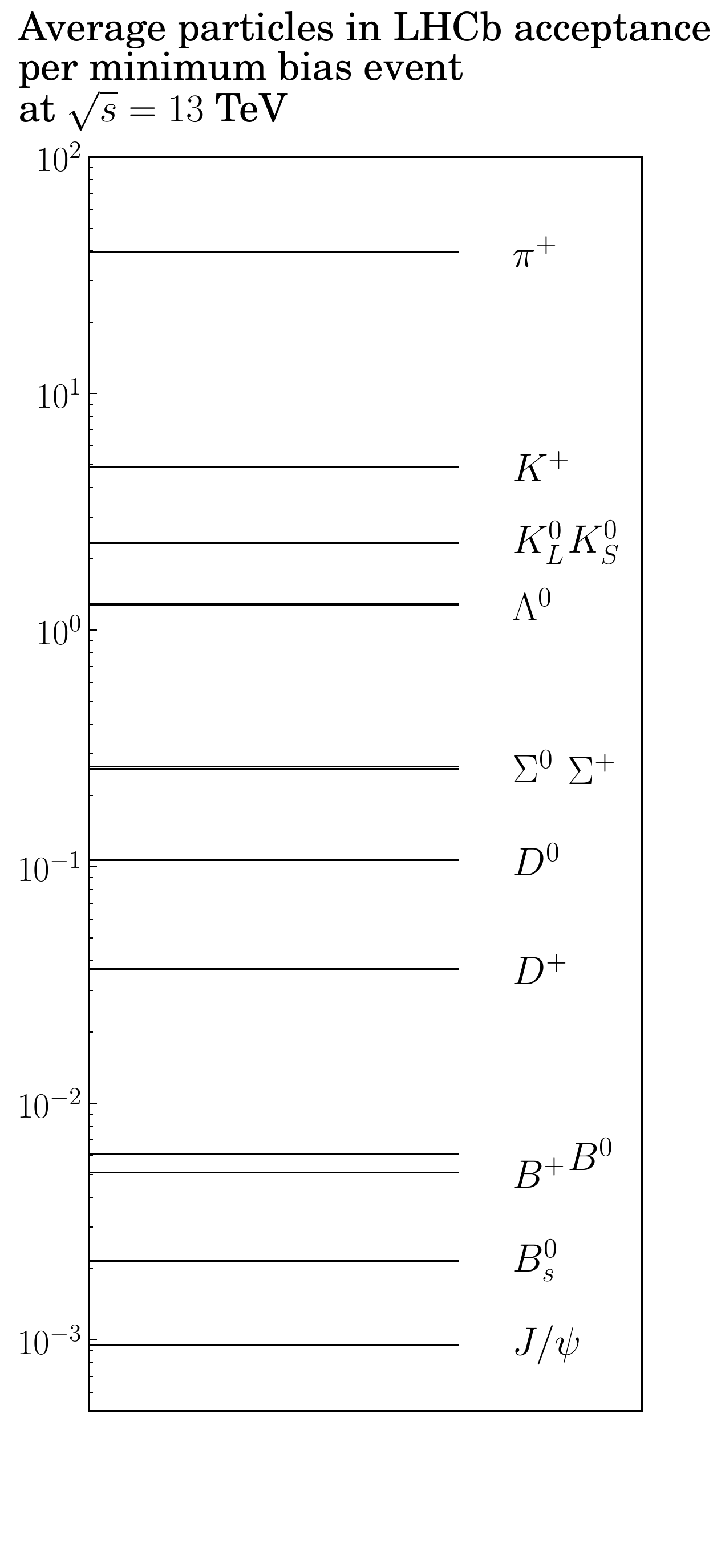}
\put(20,85){\small (c)}
\end{overpic}
\caption{In (a) decay length of strange-hadrons produced in $pp$ collisions at $\sqrt{s}=13$ TeV, compared with (b) the LHCb detector scheme. 
In (c) the average production multiplicity of different hadrons in the same kind of events. Both plots are obtained with PYTHIA Monte Carlo simulations, 
and are from Ref~\cite{Alves:2018odx}.}\label{fig1}
\end{figure}

The LHCb experiment~\cite{Alves:2008zz} at the LHC is designed and optimised to study \emph{beauty} and \emph{charm} decays
but has proven to be suitable also for strange physics. 
This, despite the average decay length of $K_S^0$ and hyperons at LHC energies is of the order of one metre, 
and it is tens of metres for $K_L^0$ and $K^+$, which are essentially considered stable in LHCb (See Figure~\ref{fig1}).
Nevertheless the huge production of these hadrons at LHC, two to three orders of magnitude larger than that of heavy flavours,
makes strange-hadron physics an increasingly exiting field at LHCb~\cite{Alves:2018odx}.

The LHCb Collaboration has already performed two world best results in this realm: the search and best upper limit on the branching fraction of \ksmumu decays~\cite{Aaij:2012rt,Aaij:2017tia,LHCb-CONF-2019-002,RamosPernas}
and the evidence for the \sigmapmumu decays~\cite{DettoriSigma}. 
These two being decays containing muons, they are probably the low-hanging fruits, nevertheless demonstrating that 
$K^0_S$ and hyperons can be used at LHCb for rare decays searches.\footnote{Incidentally, note that $K^0_S$ and $\Lambda$ hyperons are also used in LHCb as final states in many $B$ and $D$ physics 
analyses, in searches for CP violation, exploiting their dominant decay modes.  }
In this contribution I present some prospects for different strange-hadron 
searches using the LHCb data, already collected in Run 1 and 2 or to be collected in the future. 

%% Reconstruction and trigger 
The reconstruction of strange-hadrons in LHCb is limited by the tracking of their decay products
and by the trigger. 
Tracking a charged particle in LHCb usually requires to have hits in all the tracking stations (\emph{long track})
or in all but the VELO (See Fig.~\ref{fig1}) tracking stations (\emph{downstream track}). 
The former limits the decay length to 1 metre while the second allows up to 2 metres. 
Decays further downstream, albeit reconstructible,  would not allow a momentum measurement of the tracks. 
In addition to this, the trigger imposes constraints to the transverse momenta of 
the particles in an event. 
In particular during Run 1 and Run 2, typical thresholds at the level of $\pt > 1 - 5 \gevc$ were required 
for at least a muon or a hadron in the event in the first hardware level of the LHCb trigger (L0). 
Strange-hadrons are often too soft for these L0 thresholds, nevertheless given the high production rates it is sufficient that a small 
fraction of the events are triggered by the signal in question (Triggered On Signal, TOS), and in addition one can exploit event triggered by 
other particles in the same event (Triggered Independently of Signal, TIS).
On top of L0, LHCb has two software trigger layers, HLT1 and HLT2, which perform a coarse and a more refined event reconstruction, 
and select interesting events for subsequent offline analysis. 
These levels are more customisable than the L0, albeit having to respect tight rate requirements. 
During Run 1 (2010-2012) no dedicated trigger was present for strange physics at HLT1 and only 
for part of it a dedicated line at HLT2 was added for \ksmumu decays. 
In preparation for Run 2 (2015-2018), different inclusive and exclusive lines 
were studied and included in both HLT1 and HLT2~\cite{Dettori:2297352}
allowing about an order of magnitude increase in trigger efficiency for strange decays.

The data collected so far by LHCb correspond to about 10~\invfb. 
About 50~\invfb are expected to be collected in the LHCb Upgrade phase from 2021 on, 
and there is interest to continue the experiment at high luminosity with a future Upgrade, 
possibly reaching 300~\invfb~\cite{Aaij:2636441}. 
This will allow unprecedented statistics to be collected for heavy-flavours; 
while it will be represent a challenge to keep triggering also strange physics,
the highly customisable software trigger from the Upgrade phase on will allow to target with exclusive lines 
all the interesting decays, reaching efficiencies which could be up to 100\% evaluated on top of offline selected signals. 
Therefore prospects for different analyses have been prepared taking into account these possible luminosities and efficiencies, 
and some of these will be presented in the following. 

One of the most interesting decays in the short term, will be the \kspizmumu decay. 
Its $K_L$ analogous,  \klpizmumu is very sensitive to physics beyond the SM, for example it has been studied in models with extra-dimensions~\cite{Bauer:2009cf}.
However the SM prediction still suffers of a large uncertainty yielding $\mathcal{B}_{SM} (\klpizmumu) = \{ 1.4 \pm 0.3 , 0.9 \pm 0.2 \}\times 10^{-11}$. 
This uncertainty stems from a limited knowledge of Chiral Perturbation Theory parameter $|a_S|$. 
Luckily, this parameter can be extracted from a measurement of the \kspizmumu branching fraction. 
The latter is currently known with about 50\% uncertainty from a measurement by the NA48 Collaboration, 
$\mathcal{B}(\kspizmumu) = (2.9^{+1.5}_{-1.2} \pm 0.2 )\times 10^{-9}$~\cite{Batley:2004wg}. 
Any new measurement improving this branching fraction would reflect in the \klpizmumu sensitivity. 
The sensitivity to \kspizmumu decays at LHCb has been studied in Ref.~\cite{Chobanova:2195218}, with two different reconstruction methods: 
one with a full reconstruction and one without reconstructing the soft \pizero, which poses difficulties in the LHCb detector. 
This partial reconstruction still allows a good invariant mass reconstruction 
assuming a dummy \pizero, given the very low $q$ value of the decay. 
Signal efficiency is studied with Monte Carlo simulations, while backgrounds are studied in real data.
It is found that, depending on the trigger efficiency, LHCb can be competitive and improve on the NA48 measurement with about 10~\invfb of Upgrade data. 

A second group of decays which will become gradually more promising are 4-body leptonic decays. 
Again, it is the \kl mode to be the most interesting: the short distance contribution to \kfourl decays is sensitive to new physics, 
but, as often happens in strange physics, it is dominated by the long distance contribution uncertainty. 
However a measurement of the interference of $\mathcal{A} (\ksfourl)$  and  $\mathcal{A} (\klfourl)$ would give a measurement of the sign of 
$\mathcal{A} (\kl\to \gamma\gamma)$ which is a stringent test of CKM~\cite{Isidori:2003ts,DAmbrosio:2013qmd}.
While the \klfourl have been studied by different experiments,  no experimental constraints are present on  the \ks modes. 
Therefore despite the rates are expected to be very low ($\mathcal B(\kseeee) \sim 10^{-10}$ $\mathcal B(\ksmumuee) \sim 10^{-11}$, $\mathcal B(\ksmumumumu) \sim 10^{-14}$)
any limit approaching the SM rates would be a test of new physics. 

As a proxy for these decays, the \kspipiee has been studied at LHCb~\cite{MarinBenito:2193358}. 
If one considers Run 1 conditions, as expected, the efficiency is limited tightly by the L0 trigger. 
Nevertheless, given the high branching fraction, $\mathcal{ B} (\kspipiee) = (4.79 \pm 0.15 )\times 10^{-5}$, 
a yield of a hundred signal events per \invfb could be expected, with a very simple selection. 
Dedicated HLT2 trigger lines have been deployed in Run 2 and will allow for a first proof of concept. 
Upgrade trigger will instead improve the efficiency on this and related decays sensibly. 
In the ideal scenario of $\sim100\% $ trigger efficiency, with respect to offline selected events, 
of the order of $5 \cdot 10^4$ per \invfb  \kspipiee decays could be observed. 
Since one expects similar or higher efficiencies for the \ksfourl rare decays 
single event sensitivities of order $9.6\cdot10^{-10}$ over each \invfb in Upgrade conditions are expected. 

More comprehensive studies, albeit less detailed than the ones presented above,  have been done in Ref.~\cite{Alves:2018odx}, using approximate simulations of the LHCb detector. 
Dozens of decays have been studied between \ks, and hyperons showing that LHCb 
will be in the position to give many different contributions to strange-hadron physics in the near future. 
Interestingly, one can also use charged kaons at LHCb. 
While their decay length is very long compared to LHCb acceptance, it has already been shown~\cite{LHCb-PAPER-2017-049}
that high yield decays such as \kpipipi decays, can be reconstructed and used for physics calibration. 
This same decay mode has been proposed for the measurement of the $K^+$ mass, that suffers a long-standing discrepancy 
among the most precise measurements~\cite{Tanabashi:2018oca}.
In addition, with the full software trigger of the Upgrade it will be possible to study $K^+$ rare decays, 
such as \kpimumu, albeit not being competitive with the NA62 experiment. 
Finally, a field in which LHCb will also be able to contribute, are the lepton-flavour-violating decays. 
Clearly these are extremely interesting being a null test of the SM and not suffering of its uncertainties. 
As such, they have been studied by several experiments and the current upper limits read: 
$\mathcal B(K_L \to e^\pm \mu^\mp) < 4.7 \times 10^{-12}$~\cite{Ambrose:1998us}, $\mathcal B(K_L \to \pi^0 e^\pm \mu^\mp) < 7.6 \times 10^{-11}$~\cite{Abouzaid:2007aa}, 
$\mathcal B(K^+ \to \pi^+ e^- \mu^+) < 1.3 \times 10^{-11}$~\cite{Sher:2005sp}, $\mathcal B(K^+ \to \pi^+ e^+ \mu^-) < 5.2 \times 10^{-10}$~\cite{Appel:2000tc}. 
However, using $B$-physics lepton (non) universality constraints one can predict branching fractions of order $10^{-13}$~\cite{Borsato:2018tcz}. 
LHCb will be possibly able to reach this level exploiting the full Upgrade dataset~\cite{Borsato:2018tcz, Alves:2018odx}.

En passant I would like to mention also the possibilities for hyperons at LHCb. 
Concerning $\Sigma^+$ baryons, besides the \sigmapmumu decay, LHCb could improve the $\Sigma^+ \to p \gamma$ decay and try to access the $\Sigma^+ \to p e^+ e^-$ decay.
On $\Lambda$ hyperons, the branching fraction of the radiative $\Lambda\to p \pi \gamma$ could also be improved and possibly access the $\Lambda \to p\pi e^+ e^-$ decay. 
A large number of baryon number or lepton flavour violating decays have been recently constrained by the CLAS collaboration~\cite{McCracken:2015coa}: these could also be tested at LHCb. 
Higher strange-number baryons could also be probed at LHCb in order to test $\Delta S= 2$ processes, 
such as $\Xi^0 \to p \pi$ and $\Omega \to \Lambda \pi$, possibly improving limits by orders of magnitude.

In summary, the LHCb experiment has demonstrated the capability to study rare decays of \ks and hyperons, and has the possibility 
to improve many new physics sensitive decays as well as help reduce uncertainties on decays needed for \kl physics. 
With the Upgrade, the flexibility of the new software trigger will allow unprecedented rates for different decays, 
possibly establishing LHCb as a strange factory. I would like to underline here, that the decay modes mentioned in this contribution
are just examples of the possibilities and benchmarks; however new ideas area always welcome to exploit the full potential of these data.

\section*{References}
\bibliographystyle{unsrt}
\bibliography{main}
% \begin{thebibliography}{9}
% \bibitem{iopartnum} IOP Publishing is to grateful Mark A Caprio, Center for Theoretical Physics, Yale University, for permission to include the {\tt iopart-num} \BibTeX package (version 2.0, December 21, 2006) with  this documentation. Updates and new releases of {\tt iopart-num} can be found on \verb"www.ctan.org" (CTAN). 
% \end{thebibliography}

\end{document}